\documentclass[a4paper]{jpconf}
\usepackage{fancyhdr}
\begin{document}

\title{Immirzi parameter and Noether charges in first order gravity}
\author{Remigiusz Durka}
\address{Institute for Theoretical Physics, University of Wroc\l{}aw, Pl. Maksa Borna 9, 50-204 Wroc\l{}aw, Poland}
\ead{rdurka@ift.uni.wroc.pl}

\begin{abstract}

The framework of $SO(3,2)$ constrained BF theory applied to gravity
makes it possible to generalize formulas for gravitational diffeomorphic Noether charges (mass, angular momentum, and entropy). It extends Wald's approach to the case of first order gravity with a negative cosmological constant, the Holst modification and the topological terms (Nieh-Yan, Euler, and Pontryagin). 

Topological invariants play essential role contributing to the boundary terms in the regularization scheme for the asymptotically AdS spacetimes, so that the differentiability of the action is automatically secured. Intriguingly, it turns out that the black hole thermodynamics does not depend on the Immirzi parameter for the AdS--Schwarzschild, AdS--Kerr, and topological black holes, whereas a nontrivial modification appears for the AdS--Taub--NUT spacetime, where it impacts not only the entropy, but also the total mass.
\end{abstract}

\section{Black hole thermodynamics}

The discovery of the laws of black hole dynamics \cite{Bardeen:1973gs} has led to uncovering a remarkable analogy between gravity and thermodynamics. This is especially clear in the case of the first law
\begin{equation}\label{e1}
 dM = \frac{\kappa}{8\pi G}\,dA+\Omega dJ \mathrm{~(black~hole~dynamics)} \qquad
 dE=TdS+dW   \mathrm{~(thermodynamics)}\,, 
\end{equation}
where on the left we have black hole mass $M$, angular momentum $J$, angular velocity $\Omega$, area of the event horizon $A$ and surface gravity $\kappa$.

The seminal works of Bekenstein \cite{Bekenstein:1973ur, Bekenstein:1974ax} and Hawking \cite{Hawking:1974sw}, relating the area of the event horizon with the entropy
\begin{equation}
   Entropy= \frac{Area}{4l_p^2}, \qquad\qquad \mathrm{where}\qquad  l_p=\sqrt{G\hbar/c^3}\,,
\end{equation}
and the surface gravity of a black hole with its temperature
\begin{equation}
    Temperature=\hbar c\,\frac{\kappa}{2\pi}\,,
\end{equation} 
have strongly suggested that this analogy might be in fact an identity. Yet, it still lacks better and deeper understanding \cite{Jacobson:2007uj, Kolekar:2010dm, Verlinde:2010hp}.

A major step in this direction has been taken by Robert Wald, who showed that gravitational quantities from (\ref{e1}) could be obtained as the Noether charges \cite{Wald:1993nt, Iyer:1995kg}. 

The entropy can be calculated in various approaches to quantum gravity. In particular, within the Loop Quantum Gravity approach, it turns out that the key ingredient, the Immirzi parameter \cite{Immirzi:1996di, Holst:1995pc}, is explicitly present in the black hole entropy formula
\begin{equation}
    S_{LQG}=\frac{\gamma_M}{\gamma}\;\frac{Area}{4G}\,,
\end{equation}
where $\gamma_M$ is a numerical parameter valued between
$0.2$ and $0.3$. This result, coming from microscopic description and counting microstates, agrees with Bekenstein's entropy only when $\gamma$ is fixed to get rid off a numerical prefactor (for recent review see \cite{Agullo:2010zz}). Such transition is poorly understood and must be further explored (see however \cite{Ghosh:2011fc}).

This paper focuses on Wald's procedure applied to first order gravity in the setting, where the Holst modification and the Immirzi parameter are present in the action we start with. The constrained $SO(3,2)$ BF theory allows us to check if Noether approach is able to reproduce the result obtained in the LQG framework. After deriving generalized formulas for the gravitational charges we will extend results of \cite{Durka:2011yv}, and investigate a few AdS spacetimes (Schwarzschild, topological black holes, Kerr, Taub--NUT) discussing outcome in the context of the Immirzi parameter. 
\section{Wald's approach and first order gravity}

Emmy Noether's theorem concerning differentiable symmetries of the action of a physical system and the resulting conservation laws, holds well deserved place in theoretical physics.

If one considers a variation of the action, then, except field equations $(f.e.)$ multiplied by variated field, a boundary term $\Theta$ arises
$$\delta \mathcal L(\varphi,\partial \varphi)=(f.e.)\cdot \delta \varphi +d\Theta \,. $$
For any diffeomorphism $\delta_\xi\varphi=L_\xi \varphi$ being generated by a smooth vector field $\xi^\mu$, we can derive a conserved Noether current
$$
    J[\xi] = \Theta[\varphi, L_\xi \varphi] - I_\xi \mathcal L\,,
$$
where $L_\xi$ denotes the Lie derivative in the direction $\xi$ and the contraction operator $I_\xi$ acting on a p-form is given by $I_\xi \alpha_p=\frac{1}{(p-1)!}\xi^\mu \alpha_{\mu\nu^1...\nu^{p-1}}dx^{\nu^1}\wedge...\wedge dx^{\nu^{p-1}}$.

Noether current is closed on shell, which allows us to write it in the terms of the Noether charge 2-form $Q$ by the relation $J = dQ $.

In Wald's approach one applies this construction to the Einstein-Hilbert Lagrangian and its diffeomorphism symmetry \cite{Wald:1993nt, Iyer:1995kg}. The generators, being Killing vectors associated with the time translations ($\xi_t=\partial_t$) and spacial rotations ($\xi_\varphi=\partial_\varphi$), by the integration at the infinity, produce $Q[\xi_t]_{\infty}$ and $Q[\xi_\varphi]_{\infty}$, which are mass and angular momentum, respectively. Charge calculated for the horizon generator $Q[\xi_t+\Omega\xi_\varphi]_{H}$ ensures the product of the horizon temperature and the entropy. 

Outcome of this formalism agrees with other methods in different frameworks, however such a procedure applied directly to the tetrad formulation of gravity, based on Einstein-Cartan action (often called the Palatini action) with a negative cosmological constant $\Lambda=-3/\ell^2$
\begin{equation}
   S=\frac{1}{ 32\pi G}\int \left(R^{ab}\wedge e^{c}\wedge e^{d}+\frac{1}{\,2\ell^2}\,  e^{a}\wedge e^{b}\wedge e^{c}\wedge e^{d}\right)\, \epsilon_{abcd},  
\end{equation}
leads to serious problems. Noether charge evaluated for AdS--Schwarzschild metric gives wrong factor before $M$ and also requires background subtraction to get rid off cosmological divergence
\begin{equation}\label{e6}
    Mass=Q[\xi_t]_\infty=\frac{1}{2}\,M+\lim_{r\rightarrow\infty}\frac{r^3}{2G\ell^2}.
\end{equation}

To deal with this apparent problem it was suggested by Aros, Contreras, Olea, Troncoso and Zanelli \cite{Aros:1999id, Aros:1999kt} to use the Euler term as the boundary term
\begin{equation}
     32\pi G\, S=\int R^{ab}\wedge e^{c}\wedge e^{d}\,\epsilon_{abcd}+\frac{1}{\,2\ell^2}\int  e^{a}\wedge e^{b}\wedge e^{c}\wedge e^{d}\, \epsilon_{abcd}+\rho\int R^{ab}\wedge R^{cd} \,\epsilon_{abcd}\,,
\end{equation}
so for arbitrary weight $\rho$ the formula (\ref{e6}) changes into
\begin{equation}
Mass=Q[\partial_t]_\infty=\frac{M}{2}\left(1+\frac{2}{\ell^2}\,\rho\right)+\lim_{r\rightarrow\infty}\frac{r^3}{2G\ell^2}\left(1-\frac{2}{\ell^2}\,\rho\right)\,.
\end{equation}
It is easy to notice, that the fixed weight $\rho=\frac{\ell^2}{2}$ cures the result. It simultaneously removes the divergence and corrects the factor before the mass value. Moreover, with the Euler term contributing to a boundary term, adding boundary condition of the AdS asymptotics at infinity
\begin{equation} \label{e8}
(R^{ab}(\omega)+ \frac1{\ell^2}\, e^a\wedge e^b)\Big|_\infty=0
\end{equation}
and fixing the connection $\delta \omega=0$ on the horizon (in order to fix a constant temperature and fulfill the zeroth law) ensures differentiability of the action 
\begin{equation}
    \delta S_{(Palatini+\Lambda+\frac{\ell^2}{2}Euler)}=\int_{\mathcal M}(f.e.)_{a}\delta e^a+\int_{\mathcal M} (f.e.)_{ab}\,\delta \omega ^{ab}+\int_{\mathcal M}d\Theta=0\,,
\end{equation}
where $(f.e.)$ are field (Einstein and torsion) equations and boundary term 
\begin{equation}
    \Theta\Big|_{\partial\mathcal{M}}
=\epsilon_{abcd}\,\delta\omega^{ab}\wedge \left(R^{cd}(\omega)+ \frac1{\ell^2}\, e^c\wedge e^d\right)\Big|_{\partial\mathcal{M}}=0.
\end{equation}

\section{From MacDowell-Mansouri gravity to BF theory}
The formulation of gravity with the action 
\begin{equation}
  32\pi G\, S_{MM}=\int \mathrm{Palatini}+\frac{1}{\,2\ell^2}\int  \mathrm{cosmological}+\frac{\ell^2}{2}\int \mathrm{Euler}  
\end{equation}
was introduced already in the late 70's by MacDowell and Mansouri \cite{MacDowell:1977jt}. This formulation of gravity based on broken symmetry of a gauge theory has many interesting features and intriguing philosophy behind it. It combines the $so(3,1)$-spin connection $\omega^{ab}$ and the tetrad $e^a$, so the two independent variables in first order gravity, into higher gauge $so(3,2)$-connection $A^{IJ}$ through
\begin{equation}\label{e2}
A^{ab}_\mu=\omega^{ab}_\mu\,,\qquad  A^{a4}_\mu=\frac{1}{\ell}e^a_\mu\,.
\end{equation}
Notice that the Lorentz part (with indices $a,b =0,1,2,3$) is embedded in the full symmetry group of the anti-de Sitter (for which $I,J = 0,1,2,3,4$). The AdS case sets the Minkowski metric to be $\eta_{IJ}=diag(-,+,+,+,-)$.  To make dimensions right a length parameter needs to be introduced and later associated with a negative cosmological constant to recover the standard Palatini action
\begin{equation}
    \frac{\Lambda}{3}=-\frac{1}{\ell^2}\,.
\end{equation} 
One can interpret (see \cite{Wise:2006sm}) the geometric construction of $so(3,2)$-connection $A^{IJ}=(\omega^{ab},e^a)$ as a way of encoding the geometry of the spacetime $\mathcal M$ by parallel transport being "rolling" the anti-de Sitter manifold along the $\mathcal M$.

\noindent The connection 1-form $A^{IJ}$ can be further used to build the curvature 2-form
\begin{equation}
    F^{IJ}=dA^{IJ}+A^{IK}\wedge A_K^{~~J}\,,
\end{equation} 
which directly splits on torsion
\begin{equation}
   F_{\mu\nu}^{a4} =\frac1{\ell}\left( \partial_\mu e_\nu^{a}  + \omega_\mu{}^a{}_b\, e_\nu^b -\partial_\nu e_\mu^{a}  - \omega_\nu{}^a{}_b\, e_\mu^b\right)=\frac1{\ell}\left( D^\omega_\mu e_\nu^{a}-D^\omega_\nu e_\mu^{a} \right)= \frac1{\ell}\, T_{\mu\nu}^a \,,
\end{equation}
and so-called AdS curvature
\begin{equation}
   F_{\mu\nu}^{ab}= R_{\mu\nu}^{ab}+ \frac1{\ell^2}\left(e_\mu^a\, e_\nu^b-e_\nu^a\, e_\mu^b \right)\, ,   
\end{equation}
containing standard curvature
\begin{equation}
    R_{\mu\nu}^{ab}=\partial_\mu\omega_\nu^{ab} - \partial_\nu\omega_\mu^{ab} +\omega_{\mu\,c}^a\, \omega_\nu^{cb}-\omega_\nu^a{}_c\, \omega_\mu^{cb}\,.
\end{equation}
The full curvature $F^{IJ}$ could not be used for constructing an action of the dynamical theory  in four dimensions. However, one gets such a formulation with the use of the Hodge star and breaking the gauge symmetry by projecting full curvature down to Lorentz indices
\begin{equation}
    F^{IJ}\quad\rightarrow\quad \hat{F}^{IJ}= F^{ab}\qquad\qquad \mathrm{where}\qquad\qquad F^{ab}=R^{ab}+\frac{1}{\ell^2}e^a\wedge e^b\,.
\end{equation} 
Then General Relativity as a gauge symmetry breaking theory emerges from the action
\begin{equation}
    S_{MM}(A)=\frac{\ell^2}{64\pi G} \int tr\big(\hat F \wedge \star \hat F\big)=\frac{\ell^2}{64\pi G} \int \left(R^{ab}+\frac{1}{\ell^2}e^a\wedge e^b\right)\wedge \left(R^{cd}+\frac{1}{\ell^2}e^c\wedge e^d\right)\epsilon_{abcd}\,.
\end{equation}
The field equations
\begin{equation}
    \left(R^{ab}\wedge e^c+\frac{1}{2\ell^2}e^a\wedge e^b\wedge e^c \right)\,\epsilon_{abcd}=0\,,\qquad\qquad T^a=0\,,
\end{equation}
are unaffected by quadratic in curvatures Euler term, because its variation leads to $D^\omega R^{ab}$, which vanishes due to the Bianchi identity.

Recently this formalism was generalized to the form of the constrained topological BF theory with (anti-)de Sitter gauge group \cite{Smolin:2003qu, Freidel:2005ak}. This work has its roots in the work of Plebanski but instead of introducing the constraints by the Lagrange multipliers we use the term built in the same fashion like the MacDowell-Mansouri action. Apart of the 1-form $so(3,2)$-valued connection $A^{IJ}$ the action is appended with the 2-from $B^{IJ}$ field
\begin{equation}
 16\pi \, S_{BF}(A,B)= \int tr \left( F\wedge B - \frac{\alpha}{4}\hat B\wedge \star \hat B \right)\,.
\end{equation}
After solving the equations coming from variations
$$\delta B_{a5}:~F^{a5}=\frac{1}{\ell}T^a=0,\qquad\qquad\qquad \delta B_{ab}:~ F^{ab}=\frac{\alpha}{2}\epsilon^{abcd} B_{cd}$$ and plugging them back into the action, with $\alpha$ equals $\frac{G\Lambda}{3}$, we achieve $S_{BF}(A,B)\equiv S_{MM}(A)$.

In the BF formulation the Macdowell-Mansouri gravity has very interesting appearance of the perturbation theory, in which General Relativity is reproduced as a first order perturbation around the topological vacuum. Symmetry breaking occurs in the last term with dimensionless coefficient proportional to extremely small parameter $\alpha=G\Lambda/3 \sim 10^{-120}$. This opens possibility of the new approach to perturbative quantum gravity as the deformation of a topological gauge theory.

To include Holst modification \cite{Holst:1995pc} we need additional term being quadratic in $B$ fields. The action proposed by Freidel and Starodubtsev \cite{Freidel:2005ak}: 
\begin{equation}
 16\pi \, S(A,B)= \int F^{IJ}\wedge B_{IJ} -\frac{\beta}{2} B^{IJ}\wedge B_{IJ} - \frac{\alpha}{4}\epsilon^{abcd4} B_{ab}\wedge B_{cd}
\end{equation}
yields the desired extension with the Immirzi parameter 
being $\gamma=\frac{\beta}{\alpha}$, and parameters $\alpha$, $\beta$ related to the gravitational and cosmological constants
$$ \alpha= \frac{G\Lambda}{3(1+\gamma^2)}\,,\qquad
\beta= \frac{\gamma G\Lambda}{3(1+\gamma^2)}\quad \mathrm{with}\quad \Lambda=-\frac3{\ell^2}\,.$$
This model has been already investigated by the author in the several different contexts, like canonical analysis \cite{Durka:2010zx}, supergravity \cite{Durka:2009pf}, and the AdS--Maxwell group of symmetries \cite{Durka:2011nf, Durka:2011gm}. Additionally, the entropy of BF theory in the context of the entropic force was discussed in \cite{KowalskiGlikman:2010ms}. Original derivation of Noether charges related to the Immirzi parameter was presented in \cite{Durka:2011yv}.
 
Through solving the field equations for the $B$ fields
\begin{equation}\label{e22}
  B^{a4} = \frac1\beta\, F^{a4}\,,
  \qquad B^{ab}
  =\frac{1}{2(\alpha^2+\beta^2)}( \beta \delta^{ab}_{cd}-\alpha \epsilon^{ab}{}_{cd})\, F^{cd}
\end{equation}
we express the resulting Lagrangian in terms of the
$so(3,1)$-connection $\omega$, and tetrad $e$ in quite compact form:
\begin{equation}
 S(\omega,e)=\frac{1}{16\pi} \int\left( \frac{1}{4}M^{abcd} F_{ab}\wedge F_{cd}-\frac{1}{\beta\ell^2} \,T^a \wedge T_a\right)\, ,
\end{equation}
where
\begin{equation}
M^{ab}{}_{cd}=\frac{\alpha}{(\alpha^2+\beta^2)}( \gamma\, \delta^{ab}_{cd}-\epsilon^{ab}_{\;\;cd}) \equiv -\frac{\ell^2}{G}( \gamma\, \delta^{ab}_{cd}-\epsilon^{ab}_{\;\;cd}) \,.   
\end{equation}
Remarkably, this action written explicitly incorporates all six possible terms of first order gravity in four dimensions, fulfilling all the necessary symmetries, and is governed only by $G, \Lambda$, and $\gamma$ 
\begin{eqnarray}
 32\pi G\, S&=&\int R^{ab}\wedge e^{c}\wedge e^{d}\,\epsilon_{abcd}+
\frac{1}{\,2\ell^2}\int  e^{a}\wedge e^{b}\wedge e^{c}\wedge e^{d}\, \epsilon_{abcd}+\frac{2}{\gamma}\int R^{ab}\wedge e_{a}\wedge e_{b}\nonumber\\
&+&\frac{\ell^2}{2}\int R^{ab}\wedge R^{cd} \,\epsilon_{abcd} -\ell^2\gamma\int R^{ab}\wedge R_{ab}+\frac{\gamma^2+1}{\gamma}\int 2\,(T^a \wedge T_a - R^{ab}\wedge e_a \wedge e_b)\,.\nonumber
\end{eqnarray}
The structure standing behind analyzed BF model turns out to be the combination of the Cartan--Einstein action with a cosmological constant term and the Holst term, accompanied by the topological Euler, Pontryagin and Nieh-Yan terms.

Differentiability of the action is naturally incorporated in the model by the field equations and the boundary conditions specified in the previous section
\begin{equation}
  \int_{\mathcal \partial M}\delta A^{IJ}\wedge B_{IJ}\qquad \to\qquad  \int_{\mathcal \partial M}\delta\omega^{ab}\wedge B_{ab}= \int_{\mathcal \partial M}\delta\omega^{ab}\wedge M_{abcd} F^{ab}=0\,.
\end{equation} 

Despite of the different signs and form of factors related to the Immirzi parameter, we have to remember that in a derivation process the equations of motion have to be solved, which forces torsion $T^a$ to vanish, so what is left from Nieh-Yan term adds directly to the Holst term. 

Then the scheme [Palatini with $\Lambda$ + $\frac{\ell^2}{2}$ Euler] is equipped by somehow analogous combination $-\gamma$[Holst $+\frac{\ell^2}{2}$ Pontryagin], which makes further extension of Wald's procedure presented earlier rather clear and straightforward.

\section{Generalized Noether charge}
The most general form of first order gravity coming from the connection $A^{IJ}$ and the formulation of $SO(3,2)$ BF theory leads to natural generalization of the charge formula
\begin{equation}
Q[\xi]=\frac{1}{16\pi}\int_{\partial \Sigma} \frac{\delta \mathcal L}{\delta F^{IJ}}\;I_\xi A^{IJ}+(f.e.)_{IJ}\, I_\xi A^{IJ}\,.    
\end{equation}
After substituting the $B$ fields according to (\ref{e22}) and solving the field equations, which makes torsion vanish, we can write down the associated charge to be
\begin{equation}\label{20}
    Q[\xi]=\frac1{16\pi}\int_{\partial \Sigma} I_\xi \omega^{cd}\,\left(\frac{1}{2} M^{ab}{}_{cd}\, F_{ab}\right)\, ,
\end{equation}
where $\partial\Sigma$ is a spatial section of the manifold. The final form originally derived in \cite{Durka:2011yv}
\begin{equation}\label{e28}
   Q[\xi]=\frac{\ell^2}{32\pi G}\int_{\partial \Sigma} I_\xi \omega_{ab}
   \left(\epsilon^{ab}_{\;\;\;cd}F^{cd}_{\theta\varphi}-2\gamma F^{ab}_{\theta\varphi}\right)d\theta\, d\varphi\,.
\end{equation}
generalizes the results of \cite{Wald:1993nt,Iyer:1995kg}, and \cite{Aros:1999kt} to the case of first order gravity with the Holst modification.

The formula (\ref{e28}), except choosing the AdS asymptotics, was derived without specifying any initial background. To check wherever presence of the Immirzi parameter is in fact possible we have to turn to the explicit AdS spacetimes. 

\section{AdS--Schwarzschild}
We begin with the standard case of a black hole in the presence of a negative cosmological constant
\begin{equation}\label{AdS-Schwarschild}
ds^2=-f(r)^2 dt^2 +f(r)^{-2} dr^2+r^2(d\theta^2+\sin^2 \theta d\varphi^2)\,,\qquad\quad f(r)^2=(1-\frac{2GM}{r}+\frac{r^2}{\ell^2})\,.    
\end{equation}
Evaluating $I_\xi \omega_{ab}$ for the timelike Killing vector immediately forces 
\begin{equation}
    Q[\partial_t]=\frac{4\ell^2}{32\pi G}\int_{\partial\Sigma} \omega^{01}_{t}\left(\epsilon_{0123}F^{23}_{\theta\varphi}-\gamma F^{}_{\theta\varphi\,01}\right)d\theta\,d\varphi\,.  
\end{equation}
For the AdS--Schwarzschild metric given in (\ref{AdS-Schwarschild}) the term $F^{01}_{\theta\varphi}$ multiplied by the Immirzi parameter is equal to zero, thus the whole modification drops out, leaving only expression for the MacDowell--Mansouri Noether charge already obtained in \cite{Aros:1999id} and \cite{ Aros:1999kt}
\begin{equation}\label{AdS-Noether}
   Q[\partial_t]=\frac{4\ell^2\epsilon_{0123}}{32\pi G}\int_{\partial \Sigma}  \left(\frac{1}{2} \frac{\partial f(r)^2}{\partial r} \right)\left(1-f(r)^2+\frac{r^2}{\ell^2}\right)\sin \theta \,d\theta \, d\varphi\,.    
\end{equation}
Evaluating charge associated with the timelike Killing vector at infinity returns right answer for the mass
\begin{equation}
    Q [\xi ]{}_\infty=\lim_{r\rightarrow\infty}\frac{1}{4\pi }
    \int_{\partial \Sigma_{\infty}}  \left( M+\frac{\ell^2 GM^2}{r^3}\right)
    \sin \theta \,d\theta\, d\varphi = M\,.
\end{equation}
To recover the entropy from (\ref{AdS-Noether}) we first need to introduce the surface gravity $\kappa$, which can be defined (as it was done in \cite{Aros:2005by}) by the rescaled Killing vectors $\xi^a=e^a_\mu\,\xi^\mu$ in the formula
\begin{equation}
    I_\xi \omega^{a}{}_{b}\, \xi^b=\kappa\,\xi^a
\end{equation}
being a first order analog of the standard definition ($\xi^\mu \nabla_\mu \xi^\nu=\kappa \,\xi^\nu $)
in a metric formulation. Straightforward calculations shows that for the AdS--Schwarzschild
\begin{equation}
\kappa=\omega^{01}_t \Big|_{r_H}= \left(\frac{1}{2} \frac{\partial f(r)^2}{\partial r}
\right)\Big|_{r_H} \qquad\qquad T=\frac{\kappa}{2\pi}\,,
\end{equation}
Therefore, at the horizon defined by $f(r_H)^2=0$, the charge (\ref{AdS-Noether}) becomes 
\begin{equation}
     Q[\xi_t]_H=\frac{\kappa \,\ell^2}{8\pi G}\left(1+\frac{r_H^2}{\ell^2}\right)\int_{\partial \Sigma_{H}}
\sin \theta \;d\theta\, d\varphi=\frac{\kappa}{2\pi}\frac{4\pi(r_H^2 +\ell^2)}{4G}\,,
\end{equation}
so the black hole entropy yields
\begin{equation}\label{entropy1}
    Entropy=\frac{Area}{4G}+\frac{4\pi\ell^2}{4G}\, .
\end{equation}
It differs from the standard form by a constant. This does not alter the fist law, because over there we are only interested in the change of quantities (for discussion of the second law see \cite{Liko:2007vi}). Similar result appears in the Lovelock gravities, where entropy gains the term proportional to the arbitrary factor before the Gauss-Bonnet term \cite{Iyer:1994ys} (here, at least we avoid problem of the negative entropy \cite{Clunan:2004tb}). Relation between the Euler characteristic and the entropy of extreme black holes was investigated in \cite{Liberati:1997sp}.

Resulting constant has no satisfactory interpretation. We cannot go any further than noticing, that this apparent drawback of Euler regularization has the exact value of the cosmological horizon entropy \cite{Gibbons:1977mu, Bousso:2002fq} for the pure de Sitter spacetime\footnote{The de Sitter space has the different asymptotic structure, in which instead of infinity we have the cosmological horizon. The corresponding to this case analysis will be presented elsewhere.}. 

\section{Topological black holes}
Let us now focus on the topological black holes \cite{Vanzo:1997gw, Cai:1998vy,Brill:1997mf}, for which the event horizons are surfaces of nontrivial topology. The geometries of pseudo-sphere, torus, and sphere are represented by $k=-1,0,1$ in the function
\begin{equation}
    f(r)^2=(k-\frac{2GM}{r}\frac{4\pi}{\Sigma_k}+\frac{r^2}{\ell^2})\,,
\end{equation}
where $\Sigma_k$ is the unit area of the horizon hypersurface coming from the surface element
\begin{equation}
   d\Sigma_k= \left\{
\begin{array}{l l}
\displaystyle 
\sinh\theta\; d\theta \, d\phi &\mathrm{for}\qquad k=-1\\
d\theta\, d\phi &\mathrm{for}\qquad k=0\\
\sin\theta\,  d\theta \, d\phi & \mathrm{for}\qquad k=1
\end{array} \right.
\end{equation}

This generalization of geometry does not change the situation concerning the Immirzi parameter. Once again, the expression for $I_\xi\omega^{ab}$ for the field $\partial_t$ forces Immirzi contribution to be of the form $F_{\varphi \theta}^{01}$, which is exactly zero. Thus, the Noether charge 
\begin{equation}
   Q[\xi]=\frac{4\ell^2\epsilon_{0123}}{32\pi G}\int_{\partial \Sigma} \left(\frac{1}{2}\frac{\partial f(r)^2}{\partial r}\right)
   \left(k-f(r)^2+\frac{r^2}{\ell^2}\right)\,d\Sigma_k
\end{equation}
gets the values at infinity and at the horizon
\begin{equation}
   Q[\xi_t]_\infty=\frac{M}{\Sigma_k}\int_{\partial \Sigma_\infty}d\Sigma_k=M\,,\qquad
Q[\xi]_H=\frac{\kappa}{2\pi}\frac{(\ell^2 k+r_H^2)}{4G}\int_{\partial \Sigma_H} d\Sigma_k\,.
\end{equation}
Because values of unit areas $\Sigma_k$ for pseudo-sphere, torus, and sphere are $4\pi, 4\pi^2, 4\pi$, respectively, the entropy of these black holes (see \cite{Olea:2005gb}) yields the form
\begin{equation}
   Entropy =\frac{Area}{4G}+ \frac{4\pi \ell^2 k}{4G}\,,
\end{equation} 
so the torus geometry exhibits lack of the shift, and pseudo-sphere can lead to negative entropies.
\section{AdS--Kerr}
The geometry of rotating AdS-Kerr black holes can be expressed by the tetrads 
\begin{eqnarray}
e^{0}&=&\frac{\sqrt{\Delta _{r}}}{ \rho }\left(dt-\frac{a}{\Xi}\sin ^{2}\theta d\varphi \right),\qquad e^{1}=\rho
\frac{dr}{\sqrt{\Delta _{r}}},\nonumber\\  
e^{2}&=&\rho \frac{d\theta }{\sqrt{\Delta _{\theta }}},\qquad e^{3}=\frac{\sqrt{\Delta
_{\theta }}}{ \rho }\sin \theta \left(\frac{(r^{2}+a^{2})}{\Xi}d\varphi -a\,dt\right), 
\end{eqnarray}
where we additionally define
$$\rho^{2}=r^{2}+a^{2}\cos ^{2}\theta\, ,\quad \Delta _{r}=(r^{2}+a^{2})\left( 1+\frac{r^{2}}{l^{2}}\right) -2MGr\,,\quad \Delta _{\theta }=1-\frac{a^{2}}{l^{2}}\cos ^{2}\theta\,,\quad\Xi =1-\frac{a^{2}}{l^{2}}\,.$$
For the Killing vector being horizon generator ($\xi = \xi_{t} + \Omega_H \xi_{\varphi}$) we need the angular velocity 
\begin{equation}\label{e43}
    \tilde \Omega =-\frac{g_{t\varphi}}{g_{\varphi\varphi}}=\frac{a\Xi\left( \Delta_\theta(r^2+a^2)-\Delta_r\right)}{ (r^2+a^2)^2\Delta_\theta-a^2\Delta_r \sin^2\theta}
\end{equation}  
evaluated at the horizon (defined by the largest zero of $\Delta_{r}$):
\begin{equation}
    \Omega_H =\frac{a\left(1-\frac{a^2}{\ell^2} \right)}{r^2_H +a^2}\,.
\end{equation}
Besides $\Omega_H$ we will later need the value of (\ref{e43}) at the infinity $\Omega_\infty=-a/l^{2}$. 

Although this time components of (\ref{e28}) corresponding to the Holst and Pontryagin modification are not just zero anymore, the whole impact related to the Immirzi parameter, once again, drops out. The extensive summation over elements of $\omega^{ab}_{t}$, $\omega^{ab}_{\varphi}$, as well as $I_{\xi} \omega^{ab} = (\omega^{ab}_{t} + \Omega_H\omega^{ab}_{\varphi})$ gives rise to complicated expressions, but at the end, the whole Immirzi contribution is canceled out by the integration at the specified boundaries. Thus, the Noether charges for the Killing vectors associated with the time and rotational invariance stay in agreement with \cite{Olea:2005gb} and read, respectively, as 
\begin{equation}
Q\left[\frac{\partial }{\partial t}\right]=\frac{M}{\Xi}\,,\qquad
Q\left[\frac{\partial
}{\partial \varphi }\right]=\frac{Ma}{\Xi ^{2}}\,.
\end{equation}
Second expression is the angular momentum $J$, but the first quantity cannot be regarded as the mass for the Kerr-AdS black hole. As it was pointed out by Gibbons, Perry and Pope \cite{Gibbons:2004ai}, it is because the Killing field $\partial _{t}$ is still rotating at radial infinity. The non-rotating timelike Killing vector is expressed by the combination $\partial _{t}-\left( a/l^{2}\right) \partial _{\varphi }$, that substituted in the charge formula gives the physical mass 
\begin{equation}
Mass=Q\left[\partial _{t}-\frac{a}{l^{2}}\partial _{\varphi }\right] =\frac{M}{
\Xi ^{2}}\,.
\end{equation}
Also the angular velocity in the first law of black hole thermodynamics $dM=TdS+\Omega \,dJ$ should be measured relative to a frame non-rotating at infinity:
\begin{equation}
    \Omega =\Omega_H-\Omega_\infty=\frac{a\left(1+\frac{r^2_H}{\ell^2} \right)}{r^2_H +a^2}\,.
\end{equation}
To complete this analysis we finally write formulas for the surface gravity and the entropy
\begin{equation}
    \kappa=\frac{r_H \left(\frac{a^2}{l^2}-\frac{a^2}{r_H^2}+\frac{3 r_H^2}{l^2}+1\right)}{2 \left(a^2+r_H^2\right)}\,,\qquad     Entropy=\frac{4\pi  \left(r^2_H+a^2\right)}{4G \left(1-\frac{a^2}{l^2}\right)}+\frac{4\pi  l^2}{4G}\,,
\end{equation}
to find out that the entropy is again exactly of the form (\ref{entropy1}).

\section{Immirzi parameter impact}

In spite of the presence of the Immirzi parameter in the generalized formula (\ref{e28}), the resulting thermodynamics analyzed so far does not contain any trace of it. Let us try to find the reason for this disappearance by coming back to the Noether charge for our action, but now written strictly in the spacetime components. For the axisymmetric stationary spacetime with the Killing vector $\partial_\chi$, being either $\partial_t$ or $\partial_\varphi$, and definition of the connection $\omega^{ab}_\chi=e^{\nu\,a}\nabla_\chi
e^b_\nu=e^{\nu\,a}\left(\partial_\chi
e^b_\nu-\Gamma^\lambda_{\;\chi\nu}e^{b}_\lambda \right)$,
\begin{equation}
    Q[\partial_\chi]=\frac{2}{32\pi G}\int_{\partial\Sigma} \left(\epsilon^{\mu\nu}_{\quad\theta\varphi}\Gamma_{\mu \chi\nu}-\gamma\,(\Gamma_{\theta \chi\varphi}-\Gamma_{\varphi \chi\theta})\right)+\frac{\ell^2}{32\pi G}\int_{\partial \Sigma} 
   \left(\epsilon^{\mu\nu \rho\sigma }R_{\rho\sigma\theta\varphi}\Gamma_{\mu \chi\nu}-2\gamma  R^{\mu\nu}_{\;\;\;\theta\varphi}\Gamma_{\mu \chi\nu}\right)
\end{equation}
Further calculations for $\partial_t$ and the Holst part alone lead to an interesting condition
\begin{equation}\label{Holst-tribute}
-\gamma\,\int_{\partial \Sigma}(\Gamma_{\theta t\varphi}-\Gamma_{\varphi t\theta})= \gamma  \int_{\partial \Sigma}\partial_\theta g_{t\varphi}\,,
\end{equation}
which strongly suggests exploring the non-diagonal metrics. Yet, for most obvious choice being the AdS--Kerr metric we cannot achieve the goal because the non-zero expressions under integrals are canceled by the integration's limits at the horizon and at infinity. 

The author of \cite{Yu:2007zze} also notices that the Holst term contributes to the entropy formula derived from the Noether charge expression (through so-called dual horizon area), but claims that its contribution always drops out for the stationary systems, and one should turn to the dynamical black holes to observe demanded influence. 

As we will see in the next section, in spite of this claim, we can find the condition (\ref{Holst-tribute}) fulfilled by the class of an exact Einstein's solution called the AdS--Taub--NUT spacetimes, which exhibits highly nontrivial modification of thermodynamics.

The same conclusion about $\gamma$ and off-diagonal metrics, in particular the Taub--NUT spacetimes, can be drawn by looking at the Holst surface term directly in the path integral formulation \cite{Liko:2011new}. 

\section{AdS--Taub--NUT}
The Taub--NUT spacetime \cite{Misner:1963fr}, introduced by Taub, Newman, Unti and Tamburino, is the generalization of Schwarzschild metric carrying the NUT charge $n$ being gravitational analog of the magnetic monopole. This generalization to the AdS black holes with a NUT charge in 3+1 dimension
$$
    e^0=f(r)\, dt+2  n f(r) \cos\theta\, d\varphi\,\quad e^1=\frac{1}{f(r)} dr\,,\quad
e^2= \sqrt{n^2+r^2} \, d\theta\,,\quad e^3= \sqrt{n^2+r^2}\, \sin\theta\,d\varphi, 
$$
translates to the metric   
\begin{equation}
 ds^2=-f(r)^2(dt+2  n \cos\theta\, d\varphi)^2+\frac{dr^2}{f(r)^2}+(n^2+r^2)\,d\Omega^2
\end{equation}
with
\begin{equation}
    f(r)^2= \frac{r^2-2 GMr-n^2+(r^4+6 n^2 r^2-3n^4)\,\ell^{-2}}{n^2+r^2}\,,
\end{equation} 
clearly restoring the AdS--Schwarzschild solution in $n\to 0$ limit. 

This metric is problematic in many ways; it was even called by Misner a "counterexample to almost anything" in General Relativity. It has no curvature singularities, but for $\theta=0$ and $\theta=\pi$ metric fails to be invertible. It gives rise to the Misner string \cite{Misner:1963fr, Mann:1999pc, Clarkson:2005mc} being the gravitational analogue of the Dirac string. To deal with this, one has to impose a periodicity condition ensuring metric regularity. This forces the relation between the horizon radius $r_H$ and charge $n$, which leads to two separate Euclidean systems called the Taub-NUT and the Taub-Bolt solutions \cite{Mann:2004mi, Fatibene:1999ys, Lee:2008yqa}. 

In this paper we will follow only straightforward and naive evaluation of the formula (\ref{e28}) in the Lorentzian regime, leaving rigorous setting and analysis for the future paper. To this end, the Noether charge calculated for the Killing field $\partial_t$ yields the form 
\begin{eqnarray}
 Q[\partial_t]&=& \frac{  f(r) f'(r)\left(n^2+r^2\right)}{2 G}+ \gamma\;\frac{  n  f(r)^2}{2 G}\nonumber\\
&+&\frac{\ell^2}{2G}\left( f(r) f'(r)\left(1+\frac{5n^2-r^2}{n^2+r^2}\,f(r)^2\right)-\frac{2rn^2f(r)^4}{(r^2+n^2)^2}\right)\nonumber\\
&+& \gamma \,\frac{\ell^2\,n}{2G}\left(-2\left(f(r)f'(r)\right){}^2+\frac{2r f(r)^2}{r^2+n^2}f(r)f'(r)+\frac{f(r)^2}{r^2+n^2}\left( 1+\frac{3n^2-r^2}{r^2+n^2}f(r)^2\right) \right)\qquad
  \end{eqnarray}
containing the contributions from Einstein-Cartan, Holst, Euler, and Pontryagin terms. 

We find that the mass obtained from it, is appended by the term affected by the Immirzi parameter
\begin{equation}
    Mass=M+\gamma\;\frac{n \left(\ell^2+4 n^2\right)}{G \ell^2}\,.
\end{equation}
At infinity the Holst term is responsible for $n(1/2+
3n^2/\ell^2)/G$, where the Pontryagin for $n(1/2+ n^2/\ell^2)/G$, with the precise cancellations of divergent terms between them. Surprisingly, there is an intriguing coincidence between $\gamma$ addition above and the mass obtained from the Taub-NUT solution \cite{Chamblin:1998pz} coming from the periodicity conditions for the Euclidean theory.

Let us now turn to the horizon defined as usual by $f(r_H)^2=0$ with the surface gravity being equal $\kappa=f'(r_H)f(r_H)$. Noether charge for the part without topological terms
\begin{equation}
    Q[\partial_t]_{Einstein+Holst}=\frac{  f(r) f'(r)\left(n^2+r^2\right)}{2 G}+\frac{ \gamma  n  f(r)^2}{2 G}
\end{equation}
allows us to observe that Holst contribution to the entropy drops by the horizon definition. Nevertheless, the parameter $\gamma$ eventually appears in the entropy due to the Pontryagin term, which adds value proportional to the surface gravity expression \cite{Kouwn:2009qb}
\begin{equation}\label{surface}
    \kappa=\left(\frac{1}{2} \frac{\partial f(r)^2}{\partial r}
\right)\Big|_{r_H}=\frac{1}{2}\left(\frac{1}{r_H}+\frac{3(n^2+r_H^2)}{\ell^2\, r_H}  \right)
\end{equation}
placed inside the bracket of the formula 
\begin{equation}
    Q[\partial_t]_H= \frac{\kappa}{2\pi} \frac{4\pi \Big(r_H^2+n^2+\ell^2 (1-2 n \gamma  \kappa )\Big)}{4 G}\,.
\end{equation}
The final outcome contains $Area=4\pi(r_h^2+n^2)$, cosmological part $4\pi\ell^2$,  and the extra term
\begin{equation}
 Entropy=\frac{Area}{4 G}+\frac{4\pi \ell^2}{4 G}- \gamma\,\frac{2\pi n\ell^2  }{G}\,\kappa\,.
\end{equation}
Notice, that we are still in Lorentzian regime and no periodicity has been imposed, so there is no condition relating temperature to $1/(8 \pi n)$. Therefore, with the help of (\ref{surface}), we can establish
\begin{equation}
 Entropy=\frac{Area}{4 G}\left(1-\gamma\,\frac{3 n}{r_H}\right)+\frac{4\pi \ell^2}{4 G}\left(1-\gamma\,\frac{ n}{r_H}\right),
\end{equation}
which is the main result of this paper. 

\section{Conclusions}
Purpose of this review served extending the analysis discussed in \cite{Durka:2011yv} of the gravitational Noether charges for the BF theory, and reporting the nonstandard appearance of the Immirzi parameter in the AdS--Taub--NUT spacetimes.

Wald's approach for first order gravity requires regularization procedure to obtain finite charges. Remedy in the form of the MacDowell-Mansouri formulation of gravity allows for straightforward BF theory generalization to the description of black hole thermodynamics with the Immirzi parameter. The generalized formula derived from the $SO(3,2)$ BF theory action, for which topological terms are essential to secure finite charges and having well defined action principle, seems to offer formally different kind of modification, than the one known from the LQG framework. Moreover, in the explicit results for the most common AdS cases (Schwarzschild, Kerr, topological black holes) we find no trace of the desired contribution, each time facing cancellations from the both Holst and Pontryagin terms. 

The aberration appears only for the NUT charged spacetimes, where the Immirzi parameter has an impact not only on the entropy but also on the total mass, which is quite important and interesting result. In the analysis presented above parameter $\gamma$ is always coupled to the NUT charge $n$, so when $n$ is going to zero we lose the whole modification. We also report that entropy is not changed by the Holst term but by the Pontryagin term we have added to the action to avoid divergent charges at infinity and have well defined variation principle. Finally we point out that these results seem to agree with those obtained using Euclidean path integrals \cite{Liko:2011new}.

Description presented above is far from being complete. It is just restricted to straight evaluation and does not properly handle the Misner string. Careful analysis in a much wider context (especially related to the discussions carried in \cite{Hawking:1998ct}, \cite{Clarkson:2002uj}, \cite{Astefanesei:2004ji}, and \cite{Holzegel:2006gn}), will be presented in the forthcoming publication.

\ack
I am very grateful to prof. J. Kowalski-Glikman and {Tom$\acute{\mbox{a}}\check{\mbox{s}}$ Liko} for valuable discussions and helpful remarks concerning this manuscript. 
\newline

\noindent The work was supported by the National Science Center grants: 2011/01/N/ST2/00409 and N202 112740, and by the European Human Capital Program.

\end{document}